# Sedimentary Models of Fossil Biomolecules: Principles And Methodological Improvements

Wan-Qian Zhao[1]* & Li-Juan Zhao[1]

[1] School of Life Sciences, Zhengzhou University, Zhengzhou, China
* Corresponding Author, email: wqzhao@zzu.edu.cn

**Abstract**

Deamination has historically been important for authenticating ancient biomolecules. However, expanding paleogenomic datasets indicate that damage patterns are more influenced by burial hydrology and microstructural context than by molecular age or ancestry. Fossils interact with their environments differently: some form closed, water-restricted compartments that preserve minimally damaged endogenous biomolecules, whereas others serve as open molecular reservoirs in which infiltrated environmental biomolecules undergo extensive deamination from repeated water exposure. Reliance on deamination alone can therefore suppress endogenous signals and complicate the interpretation of exogenous sequences. By introducing the molecular sedimentation model for fossil biomolecules, this Perspective outlines a source-tracing framework that integrates fossil microstructure, ecological reference sets, and species-specific fragments to enable more reliable molecular inference across diverse depositional environments.

**1. Why the field needs a new framework now**
Over the past two decades, the methods developed by Dr. Pääbo's team—including deamination in DNA (cytosine to uracil, observed as C→T) and deamidation in proteins (Asn/Gln deamidation)—have been widely used as primary indicators for identifying ancient biomolecules (*1,2*). This biochemical signature played an essential stabilizing role during the early development of paleogenomics. It provided a tractable and intuitive rule: biomolecules with predictable patterns of damage were likely ancient. Under the conditions in which this paradigm was validated — cold, low-porosity, microstructurally isolated fossils — deamination was sometimes indeed a reliable temporal correlate.

However, the scope of molecular paleontology has expanded dramatically. The use of high-throughput sequencing technology now extends beyond skeletal remains to include extracts from volcanic tuff (PRJNA1309836), porous coarse pottery remains (PRJNA1309465), source rocks, and petroleum (PRJNA1091869) submitted to NCBI BioProjects (*3,4*). These materials exhibit burial and storage regimes that differ fundamentally from those of the early field. Across these diverse microenvironments, it has become increasingly apparent that deamination correlates primarily with water movement and residence time, rather than with molecular age or biological ancestry. Endogenous biomolecules preserved in sealed, dehydrated, or mineral-stabilized microdomains may retain minimal deamination, whereas environmental biomolecules entering through porous or fractured structures often accumulate extensive deamination through repeated hydration cycles.

As the field broadens, the longstanding assumption that the deamination-centered method can reliably distinguish between endogenous and exogenous sequences has become a limitation. It can suppress minimally damaged endogenous biomolecules while elevating infiltrated environmental sequences that have experienced substantial water-driven degradation. This Perspective argues that the field requires a conceptual recalibration and introduces a unified framework better suited to the empirical diversity of modern paleogenomic datasets.

**2. The molecular sedimentation model: fossils as dynamic, multi-source archives**
We propose the molecular sedimentation model for fossil biomolecules, which reframes fossils not as static preservation vessels but as dynamic, multi-source molecular sedimentation systems. In this model, biomolecules enter, accumulate, stabilize, and degrade within microenvironments governed by hydrology, porosity, mineralogy, and biological activity. Molecular patterns are therefore shaped less by elapsed time than by the physical and chemical conditions that structure water availability and molecular mobility.

This model provides a unified explanation for observations increasingly reported across paleogenomic studies: low-damage endogenous fragments preserved in sealed microdomains, high-damage environmental biomolecules infiltrating porous matrices, mixed molecular inventories reflecting episodic water flow, and inconsistent correlations between deamination and molecular age among fossils from different depositional contexts.

Three mechanistic system types
To implement this model, we categorize compartments within fossils into three types of mechanistic systems based on water accessibility and molecular transport (Fig. 1). These categories are not environmental analogies but structural regimes that determine biomolecular behavior.

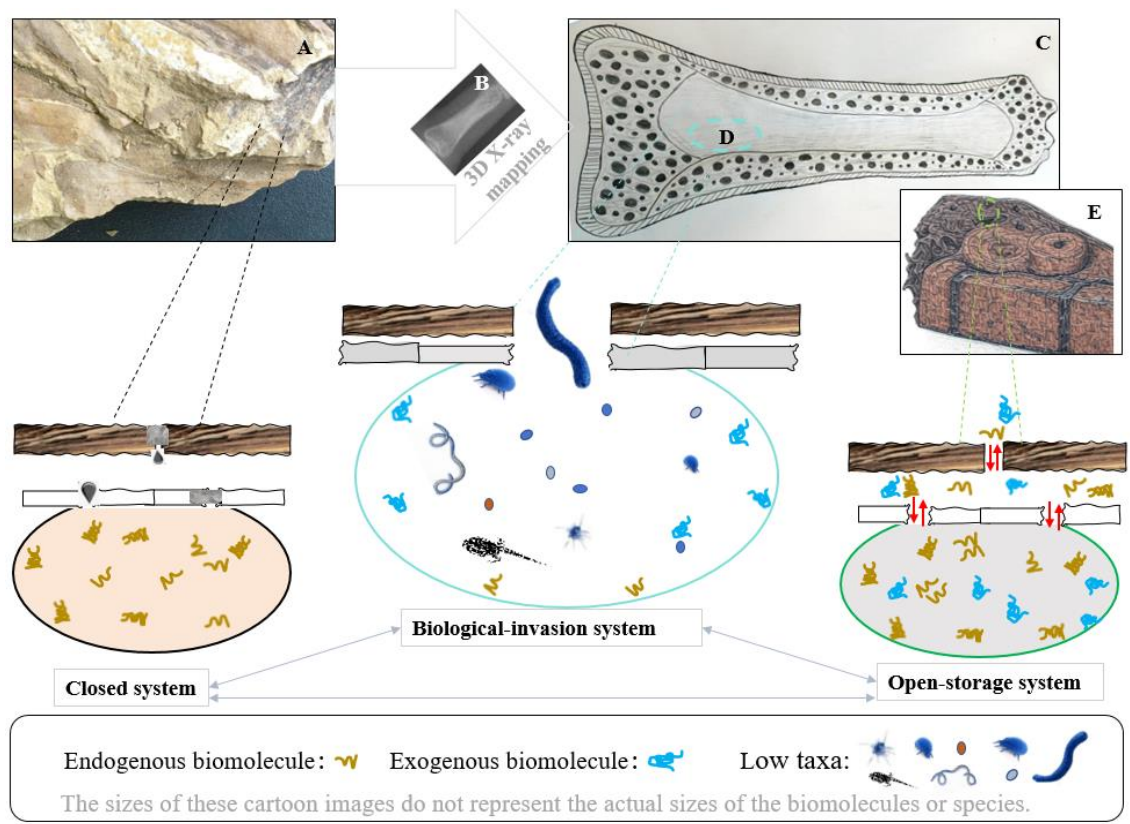

**Fig. 1. Structural contexts underlying the molecular sedimentation model**
Artistic rendering of key fossil architectures: sedimentary matrix (A), 3D X-ray reconstruction (B), long bones with internal cavities (C–D), and porous trabecular microstructure (E). These features illustrate the structural bases of the three mechanistic regimes—closed, biological-invasion, and open-storage systems.

(1) Closed systems
Microstructurally sealed or dehydrated compartments where water availability is severely restricted. Endogenous biomolecules can persist with low deamination because hydrolytic reactions are suppressed. Examples include lipid-insulated marrow cavities, salt-dehydrated zones, and mineral-occluded osteonal spaces.

Ideal microenvironments: The fossils were subjected to a short episode of heating to 55–90°C, which inactivated microorganisms and parasites, thereby suppressing endogenous decay.

(2) Biological-invasion systems
Open microspaces accessible to microbes, roots, and other biological agents. Environmental organisms introduce exogenous DNA and proteins that often dominate the molecular inventory regardless of endogenous preservation.

(3) Open-storage systems
Permeable matrices or fracture networks that permit intermittent or sustained water flow. Environmental biomolecules accumulate in stratified layers and experience substantial deamination and fragmentation through repeated hydration–dehydration cycles.

Crucially, a fossil may contain a single type or multiple types of compartments, and these compartments can transition between different states over the course of its history. This leads to significant variability in the distribution, age, and taphonomic history of its molecular content, which disrupts the straightforward relationship between storage and decay.

Across these system types, deamination reflects water activity, not chronological age. The molecular sedimentation model, therefore, shifts the central question from "Does this biomolecule show ancient damage?" to "From which microenvironment within the fossil did this biomolecule originate?"

## 3. Why is deamination alone insufficient

The mismatch between the deamination paradigm and the structure of real fossils arises from three issues:

First, deamination is a water-dependent reaction. In sealed or lipid-rich microenvironments, endogenous biomolecules can remain minimally damaged, even over extended periods. Interpreting low-damage sequences as "modern contaminants" risks discarding authentic ancient material.

Second, environmental biomolecules introduced through infiltration often experience extensive water-mediated damage, making infiltrated sequences appear superficially "ancient". This can invert the expected direction of evidence, elevating environmental signal while suppressing endogenous contributions.

Third, ultra-short fragments (<60 bp) inherently have ambiguous alignments. When deamination is applied to resolve multi-hit or low-specific ones, it creates a structural bias toward environmental sources that have undergone more extensive hydrolytic modification.

Together, these factors illustrate that deamination alone cannot serve as a reliable discriminator in the wide array of fossil microenvironments now analyzed by modern paleogenomics.

**4. A concise, actionable framework for source-tracing authentication**
Implementing the molecular sedimentation model requires shifting from single-marker authentication to multi-source provenance analysis. We outline three practical and non-disruptive steps that laboratories can adopt immediately.

(1) Species-specific fragment (SSF) targeting and source partitioning (*5*)
Utilize host-specific SSFs to measure endogenous signals. Use non-host SSFs to detect environmental sources and potential contamination. For ambiguous short fragments, integrate microstructural context rather than relying on deamination to resolve provenance.

(2) Broad-spectrum genomic reference alignment
Move beyond restricted taxonomic references. Use comprehensive genomic databases to avoid forcing environmental infiltrants into host categories. This reduces systematic misclassification, especially in porous or hydrologically active fossils.

(3) Internal-volume diagnostics before destructive sampling
Apply micro-CT, 3D X-ray mapping, and content–volume analyses to identify sealed, porous, or infiltrated microdomains. Targeted sampling increases the likelihood of capturing endogenous biomolecules while reducing environmental overprinting.

These steps transition authentication from a damage-based filter to a structured source-tracing workflow, aligning molecular analysis with the physical realities of fossil systems.

**5. Implications for the next decade of paleogenomics**
Moving to a sedimentation-based framework clarifies several field-wide issues: Sampling strategies should prioritize identifying microdomains likely to preserve endogenous biomolecules. Computational pipelines should incorporate multi-source partitioning rather than forcing deamination-driven classifications. Interpretive frameworks should treat fossils as layered molecular archives shaped by hydrological and microstructural histories rather than as uniform degradation containers.

Adopting this model allows paleogenomics to accommodate the diversity of fossils now sequenced—from warm, porous, or chemically variable environments—and reduces the risk of systematic misinterpretation.

**Conclusion**
The molecular sedimentation model provides a unified foundation for understanding why deamination patterns vary across fossil types and why their interpretive value is determined by microenvironmental context rather than molecular age. By integrating fossil microstructure, ecological reference sets, and species-specific fragments, this model establishes a more reliable and scalable framework for molecular authentication.

A sedimentation-based perspective repositions fossils not as passive containers but as dynamic molecular archives, providing a more reliable basis for the next decade of paleogenomic inference.